\documentclass[structabstract]{aa}
\usepackage{txfonts}
\usepackage{natbib}
\bibpunct{(}{)}{;}{a}{}{,}
\usepackage{epsfig}
\usepackage{graphicx}
\usepackage{subfigure}
\usepackage{latexsym}
\usepackage{pifont}
\input epsf

\newcommand{\xmm}{{\it XMM~\/}}
\newcommand{\xmmn}{{\it XMM-Newton~\/}}

\newcommand{\chandra}{{\it Chandra~\/}}

\def\Msun{\hbox{$\rm ~M_{\odot}$}}
\def\ergcms{{\rm ~erg~cm^{-2}~s^{-1}}}

\def\ergsec{{\rm ~erg~s^{-1}}}
\def\cm2{{\rm ~cm^{-2}}}
\def\chisq{{\chi^{2}}}

\def\ctsec{{\rm ~count~s^{-1}}}

\def\H0{{\rm ~km~s^{-1}~Mpc^{-1}}}

\def\eg{{\it e.g.~\/}}

\def\la{\mathrel{\hbox{\rlap{\hbox{\lower4pt\hbox{$\sim$}}}{\raise2pt\hbox{$<$}}}}}
\def\ga{\mathrel{\hbox{\rlap{\hbox{\lower4pt\hbox{$\sim$}}}{\raise2pt\hbox{$>$}}}}}
\def\d25{D$_{25}$}

\def\Ha{{H$\alpha$}}
\def\hi{H{\small I}$~$}
\def\Hii{H{\small II}$~$}
\newcommand{\Sii}{[S\,{\sc ii}]}
\newcommand{\Oiii}{[O\,{\sc iii}]}

\def\lx{L$_{\rm X}$}

\def\deg{\hbox{$^\circ$~\/}}
\def\arcm{\hbox{$^\prime$~\/}}
\def\arcs{\hbox{$^{\prime\prime}$~\/}}

\def\eps@scaling{1.0}%
\newcommand\plottwo[2]{%
  \centering
  \leavevmode
  \columnwidth=.45\columnwidth
  \includegraphics[width={\eps@scaling\columnwidth}]{#1}%
  \hfil
  \includegraphics[width={\eps@scaling\columnwidth}]{#2}%
}%

\begin{document}

\title{IKT 16: A Composite SNR in the SMC\thanks{Based on observations obtained with {\it XMM-Newton}, an ESA science mission with instruments and contributions directly funded by ESA Member States and NASA.}}

\author{R.~A.~Owen\inst{1}
\and M.~D.~Filipovi\'c\inst{2}
\and J.~Ballet\inst{1}
\and F.~Haberl\inst{3}
\and E.~J.~Crawford\inst{2}
\and J.~L.~Payne\inst{2}
\and R.~Sturm\inst{3}
\and W.~Pietsch\inst{3}
\and S.~Mereghetti\inst{4}
\and M.~Ehle\inst{5}
\and A.~Tiengo\inst{4}
\and M.~J.~Coe\inst{6}
\and D.~Hatzidimitriou\inst{7,8}
\and D.~A.~H.~Buckley\inst{9}}

\institute{Laboratoire AIM, IRFU/Service d'Astrophysique - CEA/DSM - CNRS - Universite Paris Diderot, Bat. 709, CEA-Saclay, 91191 Gif-sur-Yvette Cedex, France 
\email{richard.owen@cea.fr} 
\and 
University of Western Sydney, Locked Bag 1797, Penrith South DC, NSW 1797, Australia 
\and
Max-Planck-Institut f\"{u}r extraterrestriche Physik, Giessenbachstrasse, 85741 Garching, Germany
\and
INAF, Istituto di Astrofisica Spaziale e Fisica Cosmica Milano, via E. Bassini 15, 20133 Milano, Italy
\and
XMM-Newton Science Operations Centre, ESAC, ESA, PO Box 50727, 28080 Madrid, Spain
\and
School of Physics and Astronomy, University of Southampton, Highfield, Southampton, SO17 1BJ
\and
Department of Astrophysics, Astronomy and Mechanics, Faculty of Physics, University of Athens, Panepistimiopolis, GR15784 Zografos, Athens, Greece
\and
Foundation for Research and Technology Hellas, IESL, Greece
\and
South African Astronomical Observatory, PO Box 9, 7935 Observatory, Cape Town, South Africa
} 

\date{Received ... / Accepted ...}

\abstract 
{} 
{IKT~16 is an X-ray and radio-faint supernova remnant (SNR) in the Small Magellanic Cloud (SMC). A previous X-ray study of this SNR found a hard X-ray source near its centre. Using all available archival and proprietary \xmmn data, alongside new multi-frequency radio-continuum surveys and optical observations at H$\alpha$ and forbidden [S{\tiny II}] and [O{\tiny III}] lines, we aim to constrain the properties of the SNR and discover the nature of the hard source within.}
{We combine \xmmn datasets to produce the highest quality X-ray image of IKT 16 to date. We use this, in combination with radio and optical images, to conduct a multi-wavelength morphological analysis of the remnant. We extract separate spectra from the SNR and the bright source near its centre, and conduct spectral fitting of both regions. }
{We find IKT~16 to have a radius of $37\pm3$ pc, with the bright source located $8\pm2$ pc from the centre. This is the largest known SNR in the SMC. The large size of the remnant suggests it is likely in the Sedov-adiabatic phase of evolution. Using a Sedov model to fit the SNR spectrum, we find an electron temperature kT of $1.03\pm0.12$ keV and an age of $\approx14700$ yr. The absorption found requires the remnant to be located deep within the SMC. The bright source is fit with a power law with index $\Gamma$ = $1.58\pm0.07$, and is associated with diffuse radio emission extending towards the centre of the SNR. We argue that this source is likely to be the neutron star remnant of the supernova explosion, and infer its transverse kick velocity to be $580\pm100$ km s$^{-1}$. The X-ray and radio properties of this source strongly favour a pulsar wind nebula (PWN) origin.}
{}


\maketitle

\section{Introduction}
\label{sec:intro}


To date, $\sim275$ SNRs have been identified in the Galaxy (\citealt{green09}). The majority of these are detected as steep spectrum ($\alpha\sim-0.5$ where $S_{\nu}$=$\nu^{\alpha}$) extended radio synchrotron sources. Acceleration of the synchrotron electrons to GeV-TeV energies occurs at the shock boundary (\citealt{reynolds81}). The ambient gas is heated at the shock to X-ray emitting temperatures (\citealt{shklovskii68}; \citealt{grader70}; \citealt{chevalier77}), and radiative cooling of the gas in slower shocks produces optical emission in \Ha, \Sii~ and \Oiii~ lines (\citealt{fesen85}). Observations of the thermal X-ray and optical emission can be used to derive fundamental properties such as the age of the remnant and the local gas density. Unfortunately, the study of Galactic SNRs is impeded by two factors. The high absorption in the direction of many of these sources means that only radio emission is detected and much of the broad-band information is lost. In addition, the distance measurements to sources within the Galaxy are often uncertain by a factor of two, leading to an uncertainty of a factor of 4 in luminosity and 5.5 in supernova explosion energy (\citealt{borkowski01}). It is therefore desirable to find populations of SNRs in nearby galaxies at known distances, so that their properties can be determined with greater certainty. 

The Small Magellanic Cloud (SMC) is a very good target for study, as its 
proximity (60 kpc, \citealt{karachentsev04}) allows the 
morphology of SNRs and SNR candidates to be resolved while the low foreground Galactic 
absorption column ($6\times10^{20}$ cm$^{-2}$, \citealt{dickey90}) 
permits detection of emission in optical and soft X-ray bands. 
23 sources classified as SNRs have been discovered in the SMC 
(\citealt{filipovic05}; \citealt{payne07}). Most of these have been 
found with associated soft X-ray emission, exhibiting a variety of 
morphological structures. A synoptic study of the X-ray 
properties of thirteen of these remnants was completed by 
\citet{vdh04} (hereafter VDH04). \citet{filipovic08} 
presented \xmmn results from another three known and one candidate remnant. 
These studies found that the majority of 
SMC SNRs are likely in the Sedov-adiabatic phase of evolution.  

\begin{table*}
\caption{Details of the \xmmn observations covering IKT 16 used in this study.}
 \centering
  \begin{tabular}{ccccccccc}
\hline
Observation ID & Start Date  & Filter$^{a}$   & \multicolumn{2}{c}{Pointing co-ordinates}  & \multicolumn{3}{c}{Useful exposure (ks)}  &  Off-axis angle \\
    & (yyyy-mm-dd) & pn/MOS1/MOS2 & RA (J2000)   & Dec (J2000)  & pn   & MOS 1  &  MOS2  &  (arcmin)  \\ 
\hline
0018540101 & 2001-11-20 & M/M/M  & $00^h59^m26.8^s$ & $-72\deg09\arcm55\arcs$ &  23.4$^{b}$  & 18.1$^{b}$  & 18.0$^{b}$  & 9.3   \\
0084200101 & 2002-03-30 & T/M/M  & $00^h56^m41.7^s$ & $-72\deg20\arcm11\arcs$ &  8.8  & 10.0  & 10.3  &  8.0  \\
0110000201 & 2000-10-17 & M/M/M  & $00^h59^m26.0^s$ & $-72\deg10\arcm11\arcs$ &  14.3$^{b}$  & 16.7$^{b}$  & 16.8$^{b}$  &  9.1  \\
0212282601 & 2005-03-27 & C/M/M  & $00^h59^m26.8^s$ & $-72\deg09\arcm54\arcs$ &  0.0  & 3.8  &  3.8  &  9.3  \\
0304250401 & 2005-11-27 & M/M/M  & $00^h59^m26.8^s$ & $-72\deg09\arcm54\arcs$ &  15.9  & 17.4$^{b}$  & 17.4$^{b}$ &  9.3   \\
0304250501 & 2005-11-29 & M/M/M  & $00^h59^m26.8^s$ & $-72\deg09\arcm54\arcs$ &  14.9  & 16.5$^{b}$  & 16.6$^{b}$  &  9.3 \\
0304250601 & 2005-12-11 & M/M/M  & $00^h59^m26.8^s$ & $-72\deg09\arcm54\arcs$ &  10.6  & 16.4$^{b}$  & 16.2$^{b}$  &  9.3  \\
0500980201 & 2007-06-06 & T/M/M  & $01^h00^m00.0^s$ & $-72\deg27\arcm00\arcs$ &  14.8  & 23.2  & 23.9$^{b}$  &  11.8  \\
0601210801 & 2009-10-09 & T/M/M  & $00^h56^m15.5^s$ & $-72\deg21\arcm55\arcs$ &  23.0$^{b}$  & 24.6  & 24.6$^{b}$ &  10.4   \\
\hline    
\end{tabular}
\\
$^{a}$ - T = thin filter, M = Medium filter, C = Filter-wheel Closed.  \\
$^{b}$ - Observation used for spectral analysis.   \\
\label{table:ikt16obs}
\end{table*}

IKT 16 was discovered as a candidate supernova remnant in the SMC by 
\citet{inoue83} using {\it Einstein}, and its identity as a 
shell-type SNR was confirmed through radio and \Ha~ analysis 
(\citealt{mathewson84}). VDH04 included this remnant as part of their 
study of SMC SNRs. Using the earliest \xmmn observation, they  
found it to be X-ray faint, and its properties were hence poorly constrained. 
A region of harder X-ray emission close to the centre of the remnant, whose 
source was unclear, was also reported. 

Nine \xmmn observations 
covering IKT 16 now 
exist, giving an eightfold increase in exposure time. 
In this paper we present a study of IKT 16 and the object located towards 
its centre using X-ray data from {\it XMM-Newton}, radio data from the Australia 
Telescope Compact Array (ATCA) 
and optical images from the Magellanic Clouds Emission Line Survey (MCELS). 
In \S\ref{sec:obs}, we outline the procedures 
used to reduce and analyse \xmmn data. We also discuss complementary radio and 
optical data. In \S\ref{sec:results}, we present the results of spatial and 
spectral analysis of IKT 16 and the X-ray hard object inside it. In 
\S\ref{sec:disc}, we discuss the derived properties of IKT 16 and speculate 
on the nature of the hard source. We present our conclusions in 
\S\ref{sec:conc}.

\begin{figure*}
\centering
\includegraphics[angle=0,width=85mm]{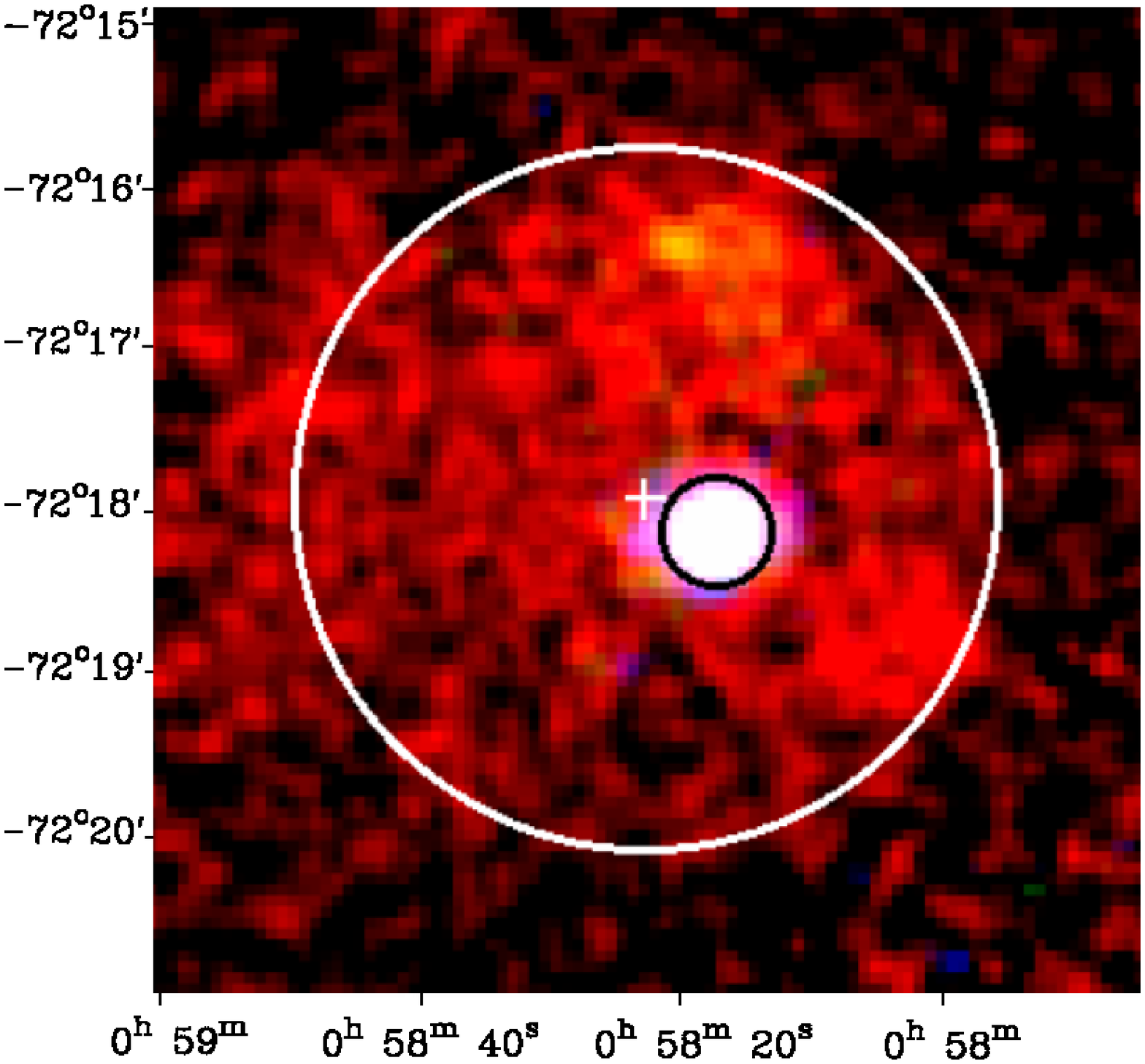}
\includegraphics[angle=0,width=91mm]{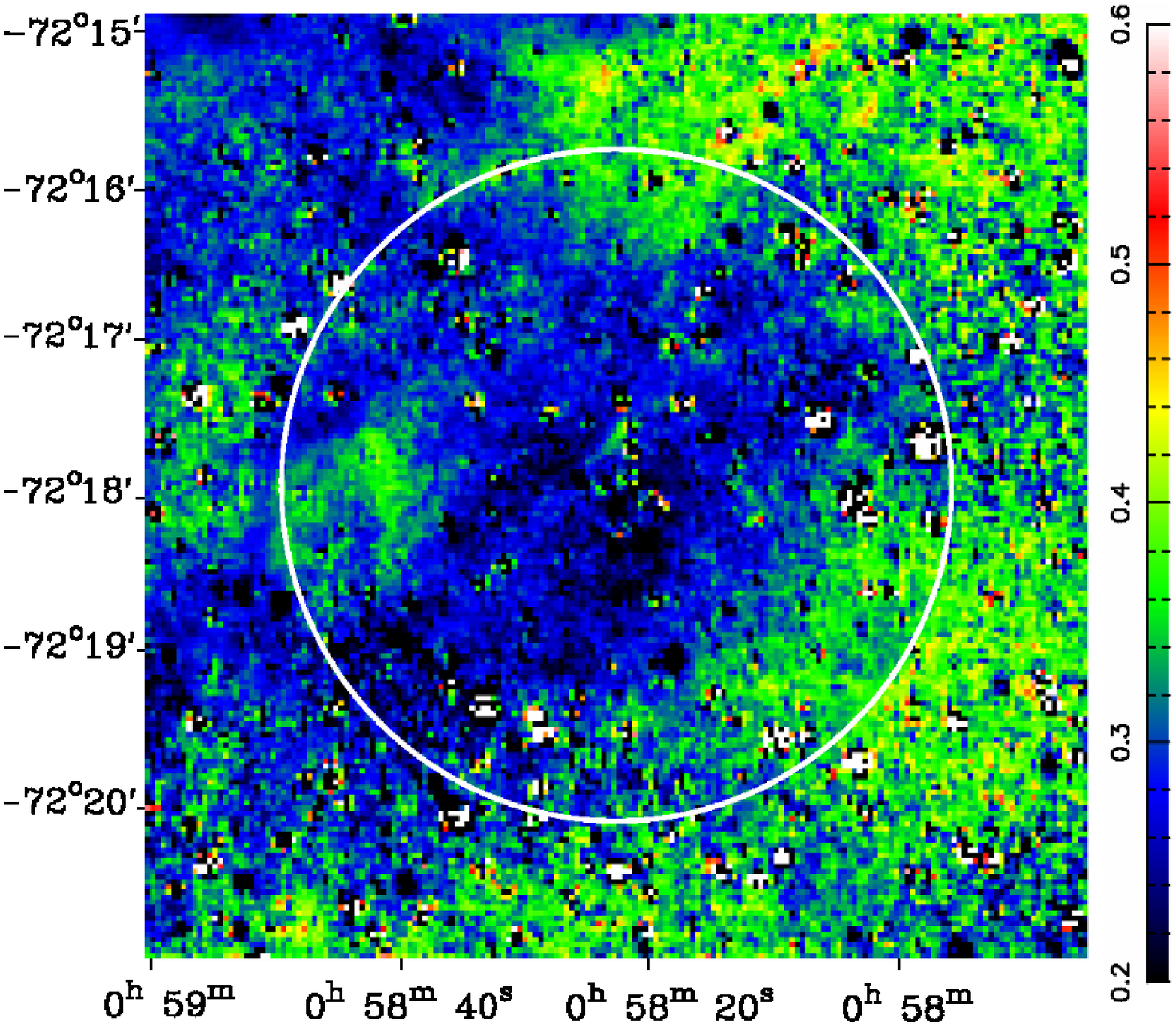}
\includegraphics[angle=0,width=85mm]{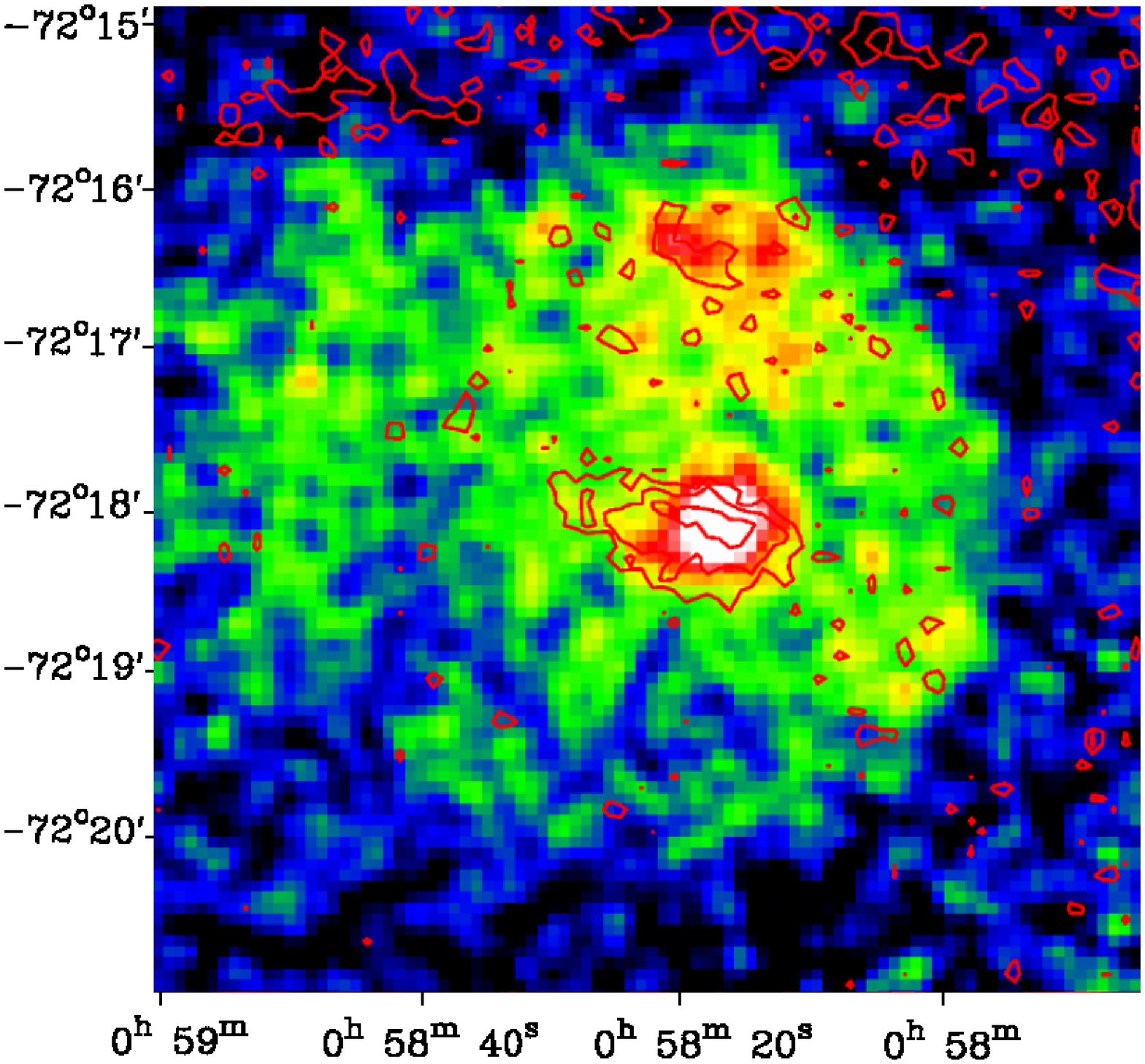}
\includegraphics[angle=0,width=91mm]{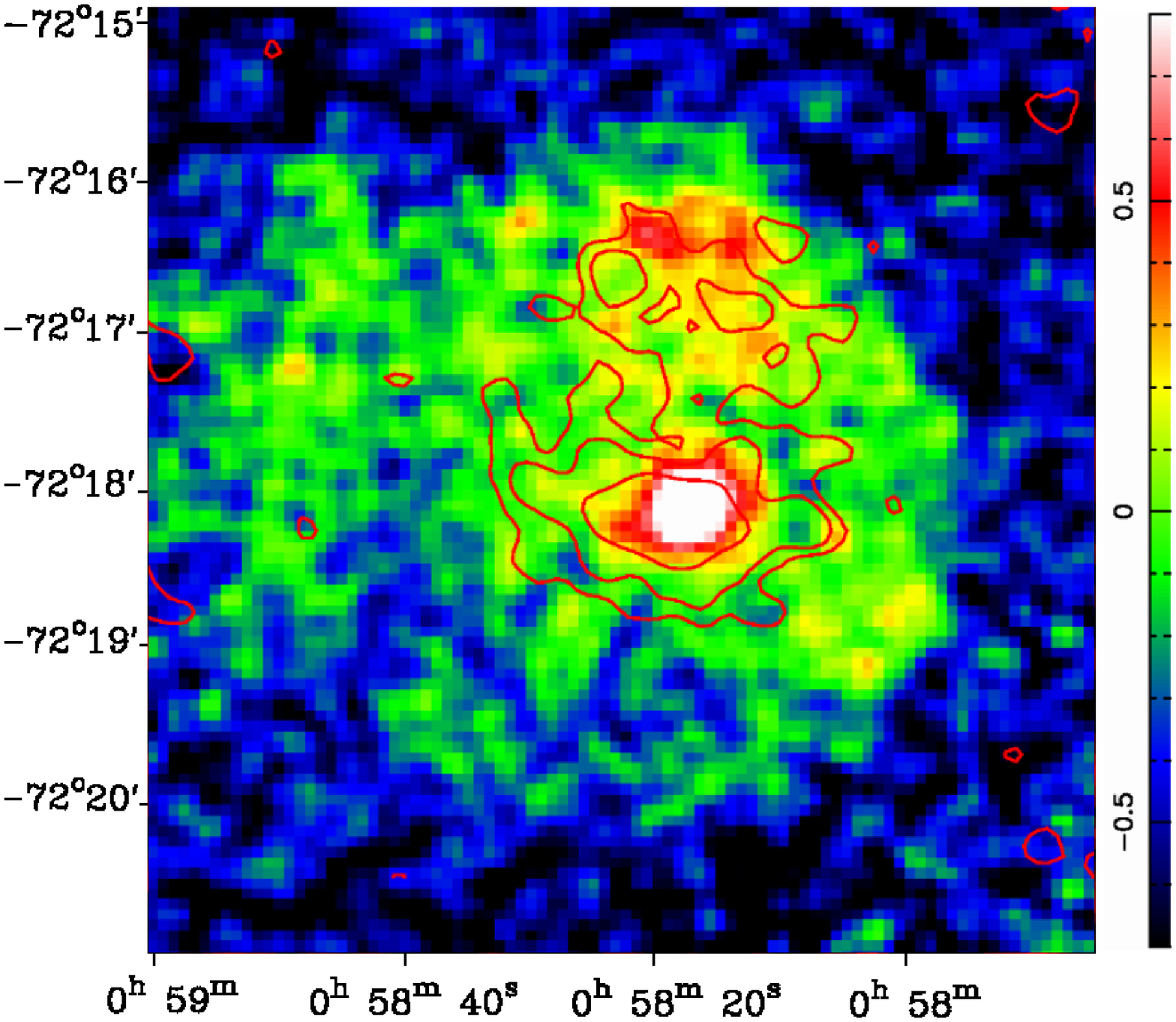}
\caption{
{\it Top left}: Lightly smoothed RGB \xmmn EPIC image of IKT~16 (R=0.3--0.8 keV, G=0.8--1.2 keV, B=1.2--2.0 keV). The two regions marked are those used for spectral analysis for the SNR (white circle) and the bright source (black circle). The centre of the remnant is marked with a cross. The image is binned in $4\arcs\times4\arcs$ pixels, and is displayed logarithmically with a dynamic range of 20. 
{\it Top right}: MCELS \Sii/H$\alpha$ ratio of IKT~16. The image is binned in $2\arcs\times2\arcs$ pixels. The white circle indicates the extent of the SNR. 
{\it Bottom left}: Lightly smoothed soft X-ray image (0.3--1 keV) overlaid with high-resolution 20~cm (1400~MHz) ATCA radio contours. The contours are at levels of 0.07, 0.14 and 0.28~mJy~beam$^{-1}$. The X-ray image is displayed logarithmically with a dynamic range of 20, and the X-ray colour bar is shown in units of log(pn+MOS) ct/pixel/20 ks.
{\it Bottom right}: Soft X-ray map overlaid with 3~cm (8640~MHz) ATCA radio contours. The contours are at levels of 0.25, 0.5 and 1.0~mJy~beam$^{-1}$. The beam sizes for the radio observations are shown in Table~\ref{table:radio}.  
}
\label{fig:ikt16images}
\end{figure*}

\section{Observations and data reduction}
\label{sec:obs}

\subsection{X-ray observations}
\label{sec:xobs}

\xmmn has observed IKT 16 serendipitously with EPIC in full-frame mode 
on 9 occasions between 2000 and 2009, at off-axis angles between 8\hbox{$^{\prime}$~\/} and 12\hbox{$^{\prime}$}. 
Eight of the nine observations were used 
for analysis, with one (obsid 0212282601) discarded due to 
persistent high background.
Details of the observations are shown in Table~\ref{table:ikt16obs}.   

Data reduction was based on SAS v10.0.0\footnote{Science Analysis Software, 
http://xmm.esac.esa.int/sas/}. 
Datasets were screened for periods of high background 
through the creation of full-field single event (PATTERN=0) lightcurves above 10 keV 
(with an upper limit of 12 keV for pn data). MOS data were excluded when the 
count-rate in a 25 second bin exceeded 0.4$\ctsec$, with pn data excluded above 
0.5$\ctsec$. These thresholds were selected based on 
quiescent levels in the observations. After filtering, useful exposure times 
range between 9 and 25 ks. 

For each observation, images and exposure maps were created in several energy 
bands: 0.3--0.8 keV, 0.8--1.2 keV and 1.2--2.0 keV for production of 3-colour 
X-ray images, and 0.3--1.0 keV for a broader-band soft X-ray image. 
We used data from single and double pixel events in the pn camera 
(patterns 0--4) and single to quadruple pixel events in MOS cameras 
(patterns 0--12). The selection FLAG=0 was used to exclude spurious data 
from bad pixels, hot columns and chip gaps, and the pixel size was set 
at 4\hbox{$^{\prime\prime}$} $\times $4\hbox{$^{\prime\prime}$}. A constant particle rate,
estimated from the corners of each image not exposed to the sky, was 
subtracted from each image. The images for each camera were then co-added 
using the SAS task {\it emosaic} in order to account for differences 
in pointing direction between the observations, and the exposure 
maps were added in the same way. Finally, the mosaic image for each camera was 
divided by the relevant exposure map and the images from the three cameras 
summed to produce a flat-field image suitable for further spatial analysis.  

Spectral analysis was performed on 15 exposures (3 pn, 5 MOS1 and 7 MOS2), 
selected on the basis of exposure time and position on the detector. When studying 
extended sources, it is necessary to be careful in defining source 
and background areas to extract. In this case there are 
two areas of interest: the central source (detected as a point source in our 
X-ray observations) located at RA(J2000)=$00^h58^m16.7^s$ Dec=$-72\deg18\arcm06\arcs$
and the remainder of the supernova remnant. The shock 
boundary of the SNR is well-described by a circular 
region centred on RA=$00^h58^m22.4^s$ Dec=$-72\deg17\arcm52\arcs$ 
with a radius of $128\pm8$\hbox{$^{\prime\prime}$}, with the error due to
uncertainty in the best fit circle radius. We therefore defined 
two source regions for each observation: a 20$\arcs$ radius area centred on 
the bright source and a 128$\arcs$ radius area centred on our best estimate 
of the centre of the SNR, excluding the previously defined source area. 
These extraction regions are shown on Fig.~\ref{fig:ikt16images} (top left). For 
the first region, a background region was extracted from a nearby position 
on the same CCD outside the SNR. For the second, a nearby background region 
of similar size 
to the source region was found such that the average exposure time for source 
and background regions was the same. This was done to ensure that the 
contributions from the cosmic X-ray background (vignetted by the telescope) 
and particle background (not vignetted to first order) remained in the correct 
proportion when applied to the source spectrum. The SAS tasks {\it arfgen} and 
{\it rmfgen} were used to generate appropriate Auxiliary Response Files (ARF) 
and Response Matrix Files (RMF) for each observation, and the counts in 
adjacent spectral bins were summed to a minimum of 20 counts per bin 
using the Ftool\footnote{http://heasarc.gsfc.nasa.gov/ftools/} {\it grppha}.

\subsection{Optical observations}
\label{sec:optobs}

The Magellanic Cloud Emission Line Survey (MCELS) was carried out using the University of Michigan/CTIO Curtis Schmidt telescope. Using narrow band filters corresponding to \Ha (6563\AA), \Sii (6724\AA) and \Oiii (5007\AA) line emission, plus red and green continuum bands (used to subtract most of the stars from the images), images of diffuse structure across both Magellanic Clouds have been produced with an angular resolution of 2.4\arcs/pixel. An image of the \Sii/\Ha~ ratio across the remnant is presented in Fig.~\ref{fig:ikt16images} (top right). Further information about the MCELS is given in \citet{smith00}. 

Integral Field Spectroscopy (IFS) was performed on 3~November, 2010 at Siding Springs Observatory using the 2.3-m Advanced Technology Telescope and its Wide Field Spectrograph (WiFeS). WiFeS provides a 25\arcs $\times$ 38\arcs field with 0.5\arcs per pixel spatial sampling along each of 25\arcs $\times$ 1\arcs slits. The output format match two 4096$\times$4096 pixel CCD detectors in each camera individually optimized for the blue and red ends of the spectrum. The 900~second single exposure was made in the central region of IKT~16 at position angle (east of north) 90 degrees under clear skies with seeing estimated at 1\arcsec. The exposure was made in classical equal mode\footnote{This is a principal WiFeS data accumulation mode in which data were accumulated in the red and blue cameras on the source for equal times.} using the RT560 beam splitter and 3000 Volume Phased Holographic (VPH) gratings. For these gratings, the blue (708~lines~mm$^{-1}$) range includes 3200--5900\AA\ while the red (398~lines~mm$^{-1}$) range includes 5300--9800\AA.

The data were reduced using the WiFeS data reduction pipeline based on NOAO (National Optical Astronomy Observatory) IRAF software. This data reduction package was developed from the Gemini IRAF package (\citealt{2003SPIE.4841.1581M}). The pipeline consists of four primary tasks: {\it wifes} to set environment parameters, {\it wftable} to convert single extension FITS file formats to Multi-Extension FiTS ones and create file lists used by subsequent steps, {\it wfcal} to process calibration frames including bias, flat-field, arc and wire; and {\it wfreduce} to apply calibration files and create data cubes for analysis. 


\subsection{Radio observations}
\label{sec:radioobs}

We have extracted all archival radio-continuum observations of IKT~16 using the Australia Telescope Compact Array (ATCA)  comprising observations at 4 radio wavelengths: 20, 13, 6 and 3~cm for ATCA (\citealt{dickel01}; \citealt{filipovic02}; \citealt{2004MNRAS.355...44P}; \citealt{1997A&AS..121..321F}). In addition we make use of the Molonglo Observatory Synthesis Telescope (MOST) SMC survey at 36~cm (843 MHz) \citep{turtle98}. The beam size for these observations ranges from 7{\hbox{$^{\prime\prime}$}} to 45{\hbox{$^{\prime\prime}$}}. IKT~16 was also observed with the ATCA as a part of projects C281 and C634. More information about observing procedure and other sources observed during these sessions can be found in \citet{bojicic07}, \citet{bojicic10} and \citet{2008SerAJ.176...59C}. Soft X-ray images of IKT~16 overlaid with 3~cm and high-resolution 20~cm radio contours are shown in Fig.~\ref{fig:ikt16images}, and the integrated flux densities detected from our radio observations of the SNR are shown in Table~\ref{table:radio}. The error in the measured integrated flux density (for the whole system and the bright source) is estimated as a quadrature sum from the local noise level (0.1~mJy~beam$^{-1}$) and the uncertainty in the gain calibration (10\%). For the SNR region (excluding the bright source) the error is calculated by adding the errors for the other two extraction regions in quadrature.

\begin{figure}[t]
\centering
\rotatebox{0}{\scalebox{0.5}{\includegraphics{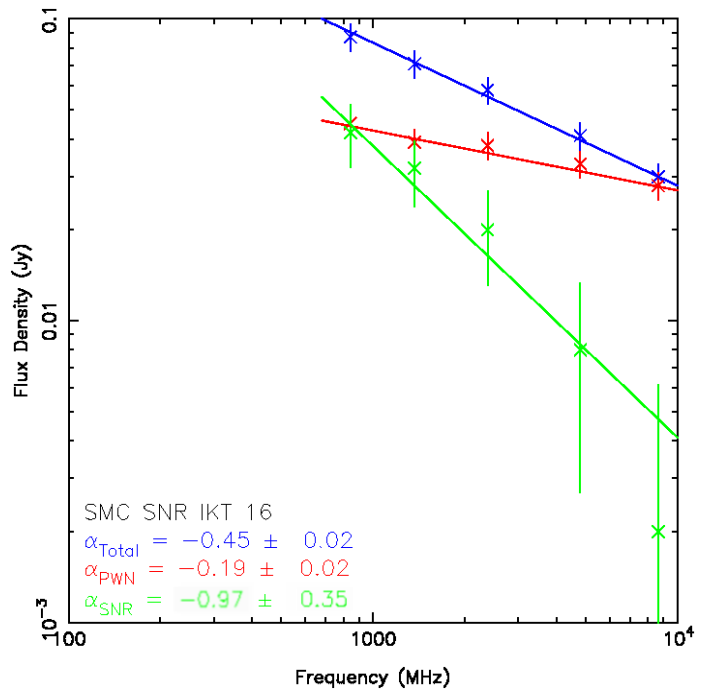}}}
\caption{Radio spectra of the entire SNR (blue), the bright source near the centre 
of the remnant (red) and the remainder of the remnant (green), constructed from 
radio observations at 5 frequencies using MOST and ATCA. The best fit 
power-law fits are shown, and the data is also shown in Table~\ref{table:radio}.
}

\label{fig:radiospec}
\end{figure}

\begin{table*}[t]
\caption{Radio Integrated Flux Density of IKT 16.}
\begin{center}
\begin{tabular}{ccccccl}
\hline
$\nu$ & $\lambda$  & Beam Size  & S$_\mathrm{Total}$ & S$_\mathrm{Source}$ & S$_\mathrm{SNR}$ & Telescope/Project \\
(MHz) & (cm)      & (arcsec) & (Jy) &(Jy)  & (Jy)  &\\
\hline
843  & 36  & 44.9$\times$43   & 0.087 & 0.045 & 0.042 & MOST\\
1371 & 20  & 7.05$\times$6.63 & 0.071 & 0.039 & 0.032 & C281\\
2378 & 13  & 7.02$\times$6.58 & 0.058 & 0.038 & 0.020 & C281\\
4798 & 6  & 30$\times$30     & 0.041 & 0.033 & 0.008 & C634, Parkes\\
8640 & 3  & 20$\times$20     & 0.030 & 0.028 & 0.002 & C634, Parkes\\
\hline
\end{tabular}
\end{center}
\label{table:radio}
\end{table*}

\section{Results}
\label{sec:results}

\subsection{X-ray}
\subsubsection{Morphology}
\label{sec:morph}

Fig.~\ref{fig:ikt16images} shows a 3-colour X-ray (0.3--0.8 keV, 0.8--1.2 keV and 1.2--2.0 keV) image of IKT~16, along with radio images from ATCA and optical data from MCELS. The X-ray images are lightly smoothed with a Gaussian kernel of width 4{\hbox{$^{\prime\prime}$}}.  We define the boundary of the SNR to be where the surface brightness of the X-ray emission abruptly falls to the background level. Using this definition, the radius of the remnant is estimated to be $128\arcs\pm8${\hbox{$^{\prime\prime}$}}, with the error resulting from a 2-pixel uncertainty in fitting a circular region to the edge of the X-ray emission. At the assumed distance of the SMC (60~kpc), this corresponds to a radius of 37$\pm$3~pc, consistent with being an older remnant in the Sedov phase of evolution (\citealt{cox72}). The measured radius makes it the largest known SNR in the SMC. The shock boundary is well-defined in the north and west of the SNR, where the diffuse X-ray emission is strongest. We find small variations in the temperature measured across the remnant, with the region of strongest X-ray emission in the north slightly hotter than its surroundings.  More tenuous emission is observed to the east of the remnant, and there is some evidence for greater extension of the SNR in this direction. Depending on the direction of measurement, the linear radius of the SNR varies between 122\arcs (in the N-S direction) and 144\arcs (NE-SW), implying that the remnant is not perfectly circular on the sky. These deviations are relatively small and our simplification of the geometry of the system does not have a large effect on the results we obtain. These variations do however lead to uncertainty in the position of the SNR centre, which limits the precision with which we can measure the dynamics of the system.    

The bright source observed inside the SNR shell is significantly harder than the emission from the rest of the remnant, as seen in the 3-colour image and its X-ray spectrum (in Section~\ref{sec:xspec}). We find no evidence for this source being extended in \xmmn measurements. We therefore limit its maximum X-ray extent to the FWHM of the detector (6{\hbox{$^{\prime\prime}$}}), which is 1.7~pc at the distance of the SMC. Assuming that the SNR shell can be described as spherically symmetric, the position of the bright source is offset from the centre of the shell by $30\arcs\pm8\arcs$ (with the error mostly due to uncertainty in the position of the SNR centre). Not taking into account possible projection effects, this corresponds to a linear distance of $8\pm2$~pc. We note that the bright source has been previously identified in a 9 ks \chandra observation (OBSID 2948; \citealt{evans10}). The \chandra X-ray position is consistent with that found here, but due to the extremely short observation time and large off-axis angle in that observation, the errors on the \xmmn position are smaller. The observation also suggests that the source may be slightly extended (with an extent of 1.1\hbox{$^{\prime\prime}$}$\pm$0.6\hbox{$^{\prime\prime}$}). A longer on-axis observation is necessary to constrain this value precisely.   

\subsubsection{Spectral analysis}
\label{sec:xspec}

Spectral analysis was performed using XSPEC v12.5.1 (\citealt{arnaud96}). For each of the relevant observations, source and background spectra were extracted as described in the previous section. All 15 spectra extracted for each of the two extraction regions were fitted simultaneously, allowing only for a renormalisation factor between observations from different cameras. To account for photoelectric absorption by interstellar gas, two absorption components were used in fitting: a {\it phabs} component fixed at the foreground Galactic value of $6\times10^{20}\ergsec$ (\citealt{dickey90}) assuming elemental abundances of \citet{wilms00}, and a variable absorption ({\it vphabs}) component to account for absorption inside the SMC. This second component has metal abundances fixed at 0.2 solar, as is typical in the SMC (\citealt{russell92}).  

X-ray emission from the central region was fit with an absorbed power-law, and emission from the remainder of the SNR was fit with a Sedov model (\citealt{borkowski01}). Due to the relatively broad point spread function of the \xmmn EPIC detector, we estimate that for a point source, $\sim$25\% of the source counts will fall outside an extraction region of $20\arcs$. Therefore, an extra power-law component is introduced to the SNR fit, with index fixed to that obtained through fitting the central region and normalization set to account for this spillover. Similarly, as the central extraction region encompasses 2.5\% of the overall SNR region, an appropriately scaled Sedov component is added to the central source model, with parameters fixed to those derived for the SNR. 

The free parameters of the Sedov model are the mean shock temperature, electron temperature just behind the shock front, metal abundances, ionization age (electron density behind the shock front multiplied by the remnant age) and normalization. Providing that the SNR has spherical symmetry and is in the Sedov phase of evolution, the fits obtained can be used to estimate several physical parameters. Using the distance to the SNR ($D$) and radius ($R$ in m, assuming a distance of 60~kpc), volume of X-ray emitting material ($V$ in m$^3$), normalization derived from the XSPEC fit ($N$), shock temperature ($T_S$ in keV) and baryon number per hydrogen atom ($r_m\approx1.4$, assuming a helium/hydrogen ratio of 0.1), it is possible to derive the electron density ($n_e$), age of the remnant ($t_{dyn}$), total emitting mass ($M$), initial explosion energy ($E_0$) and ionization age ($I_t$) through the following system of equations (VDH04, from \citealt{borkowski01}):     

\begin{equation} 
\label{eq:ne}
n_e = \sqrt{\frac{3D^2N}{10^{-24}R^3}}, ~$m$^{-3} 
\end{equation}

\begin{equation}
\label{eq:age}
t_{dyn} = \frac{1.3\times10^{-14}R}{\sqrt{T_S}}, ~$yr$ 
\end{equation}

\begin{equation} 
\label{eq:mass}
M = 5\times10^{-31}m_pr_mn_eV, \Msun 
\end{equation}

\begin{equation}
\label{eq:energy} 
E_0 = 2.64 \times 10^{-8} T_sR^3n_e, ~$erg$ 
\end{equation}

\begin{equation} 
\label{eq:ion}
I_t = 4\times10^{-6} n_et_{dyn}, ~$m$^{-3} $s$ 
\end{equation}

\begin{figure*}
\centering
\rotatebox{270}{\scalebox{0.4}{\includegraphics{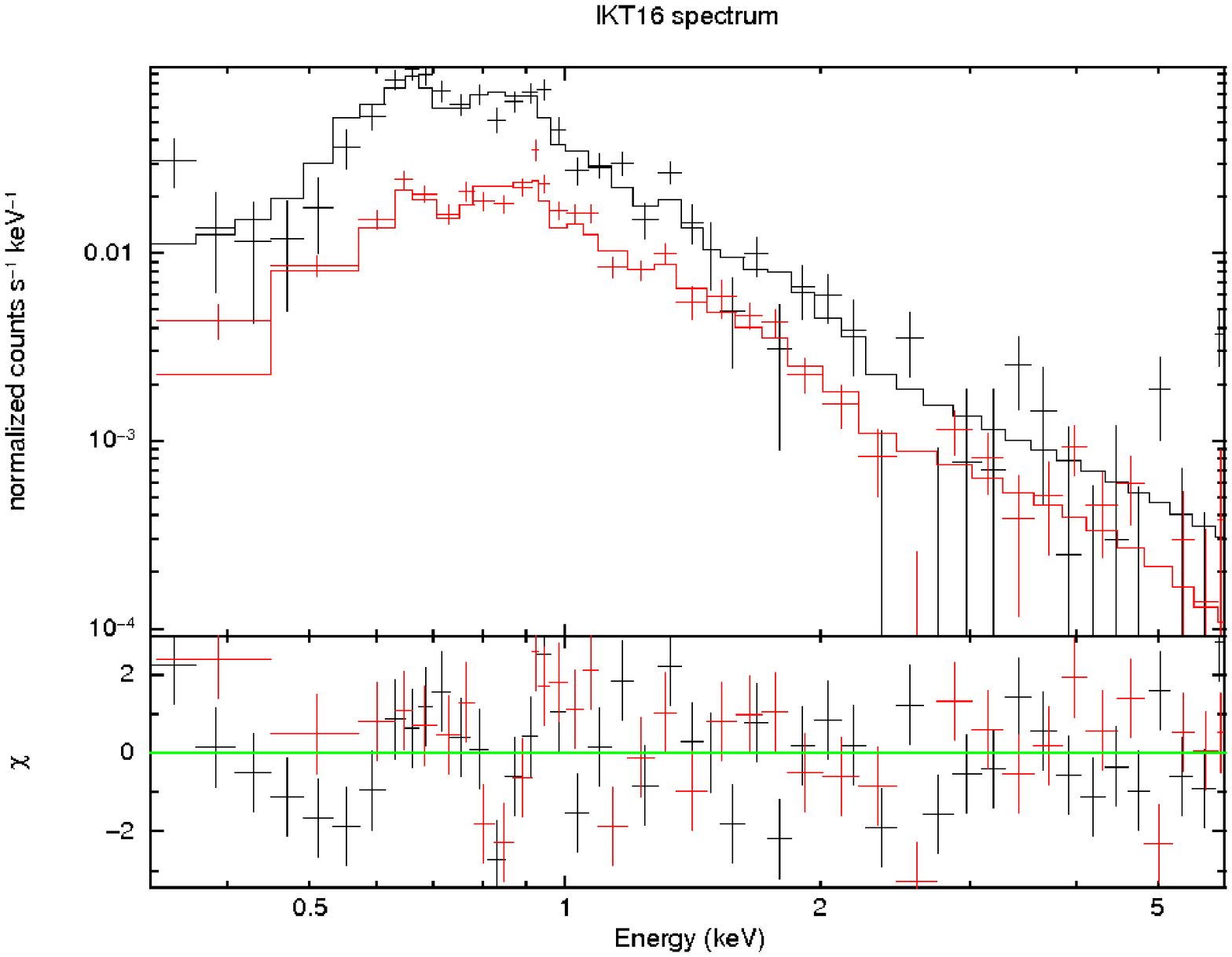}}}
\rotatebox{270}{\scalebox{0.44}{\includegraphics{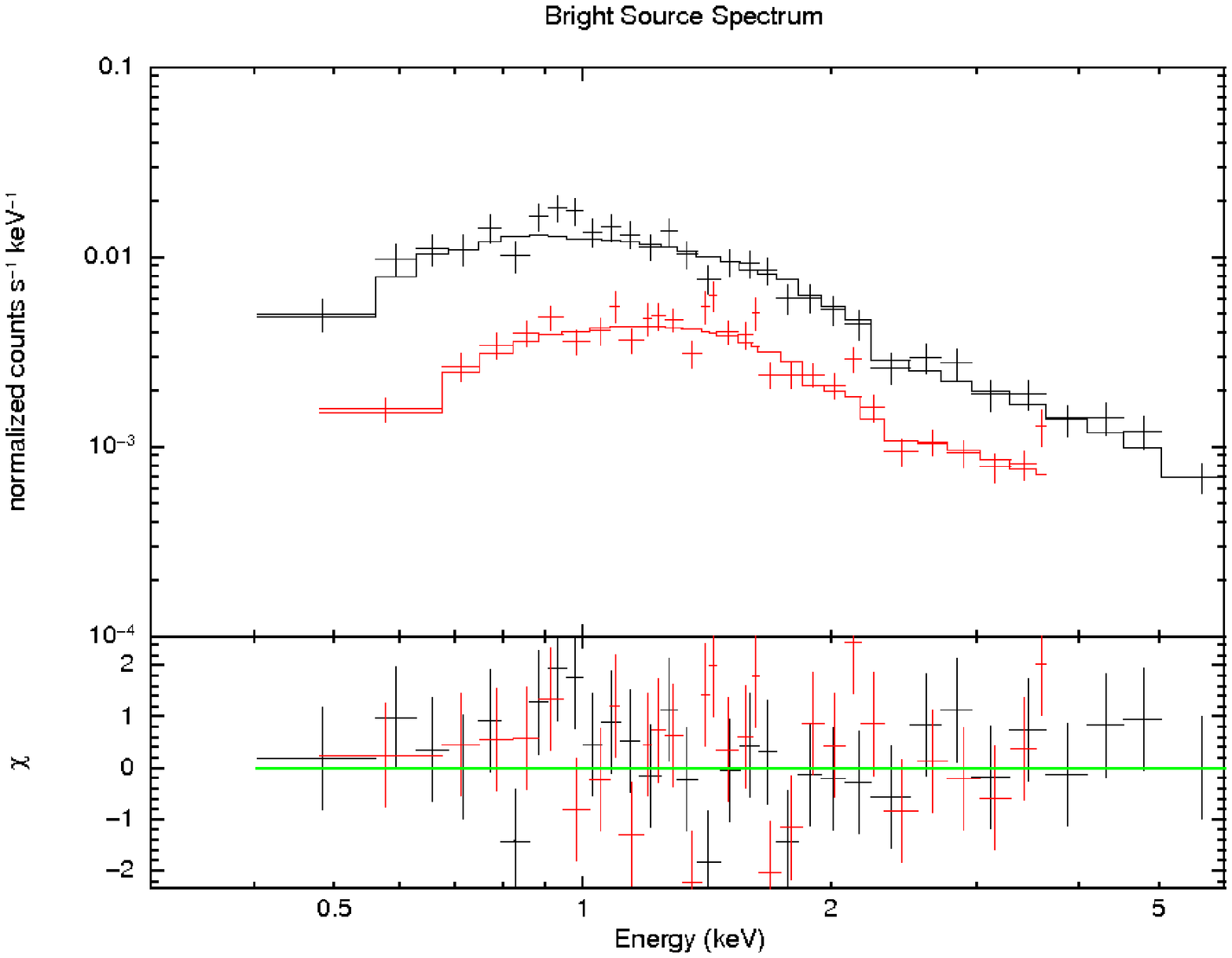}}}
\caption{EPIC spectra from IKT 16 (left) and the bright source located near its centre (right). In both cases the solid line corresponds to the best-fit spectral model (see text and Table~\ref{table:specfit}), with the upper line showing the combined spectra from pn observations and the lower line combined spectra from MOS1 and MOS2 observations. The $\chi^{2}$ residuals with respect to the best-fitting model are also shown. }
 \label{fig:xrayspec}
\end{figure*}


\begin{table*}
\caption{Parameters of the best-fit model to SNR and bright source regions.}
\centering
\begin{tabular}{lccccccc}
\hline

Region  & SMC Absorption &  Power-law Cont. &  \multicolumn{2}{c}{Sedov model}  & Goodness  &  Unabsorbed \lx$^{a}$ \\
& $10^{21}$cm$^{-2}$  &  Photon Index  &   Shock T (keV)  &  Ion. time ($10^{10}$cm$^{-3}$s)  & of Fit &  0.5--10 keV  \\
& &  Normalization  &  Normalization  &   &  $\chi^{2}$/dof  & ($10^{35}\ergsec$) \\
\hline
SNR  & 3.4(f)  &  1.58(f)  &  1.03$\pm$0.12  & 6.1 & 1103/1032  &  1.6$\pm$0.4  \\
&  &  $1.0\times 10^{-5}$(f)  & $1.4\pm0.3\times 10^{-4}$   &    &    &    &    \\
\\
Bright Source  &  3.4$\pm$0.6  &  1.58$\pm$0.07  &  1.03(f)  &  6.1(f) & 150/137  & 1.6$\pm$0.2   \\
&  &   $3.4\pm0.4\times 10^{-5}$  &  $3.6\times 10^{-6}$(f)  &   &    &    \\
\\
\hline    
\end{tabular}
\\
$^{a}$ - Includes correction for spillover of bright source photons into SNR extraction area and vice versa. \\
(f) - Parameter fixed for consistency between fit regions. \\ 
\label{table:specfit}
\end{table*}


We note that our distance estimate of 60 kpc to IKT 16 may not be entirely accurate due to the large line of sight extent of the SMC, which is measured to be 4-6 kpc in this region of the galaxy (\citealt{subramanian09}). Errors in the distance measurement will also impact the measured radius of the SNR ($\propto$ D) and its volume ($\propto$ D$^3$), and the physical properties derived from the above system of equations will be affected.   

Details of the best fit spectra obtained are given in Table~\ref{table:specfit} and the combined pn and MOS spectra for both extraction regions are shown in Fig.~\ref{fig:xrayspec}. Due to the limited quality of the spectral data, shock and electron temperatures are assumed to be the same (valid as an evolved SNR should be approaching electron-ion equilibration). The metal abundances in the fits are fixed at 0.2 solar for the same reason. Since lines, whose presence is required to fit the ionization age, are not easily discernable in the X-ray spectrum, this parameter was calculated based on the above system of equations. Spectral fitting was then repeated with the calculated ionization age to derive new Sedov parameters, and this process was repeated until convergence was achieved. The best fit physical parameters from the Sedov model fit are shown in Table~\ref{table:sedov}. The best fit models found require a significant absorption component within the SMC in addition to foreground Galactic absorption. We initially left this absorption as a free parameter for both regions. In this case, we find the best fit Sedov model for the SNR to have an electron temperature of 0.71~keV with absorption inside the SMC of $4.6\times10^{21}$ cm$^{-2}$ ($\chisq=1100.7/1032$), outside the error bounds of the absorption found in the bright source spectrum. Setting the SNR absorption at the level found for the bright source does not significantly worsen the fit ($\chisq=1103.2/1033$), but it does alter the best fit temperature (see Table~\ref{table:specfit}). We fix the absorption in the SNR to the level of the bright source, as is expected if the two emission regions are physically connected, and derive the Sedov parameters for the SNR based on this fit (see Table~\ref{table:sedov}). If we instead fix the absorption for the bright source at the level found for the SNR, we obtain a significantly worse fit for this component ($\chisq = 164/137$). In this case, we find the power-law index of the source to be somewhat steeper ($\Gamma = 1.74\pm0.08$). For both components, the average flux across all observations is measured to be $2.7\times10^{-13}\ergcms$ (0.5--10 keV), corresponding to an unabsorbed luminosity of $1.6\times10^{35}\ergsec$. The flux for both components is found to be constant between observations to within $15\%$. There is thus no evidence for any long-term variability of the point source. 

The X-ray spectrum of the bright source within the SNR is typical of many astrophysical sources. The possible nature of this source will be discussed in Section~\ref{sec:discsrc}.

\subsection{Optical}
\label{sec:optical}


Using data from MCELS, we find that the \Sii/H$\alpha$ ratio found in the SNR is $\sim0.25$ (Fig.~\ref{fig:ikt16images}, top right), significantly below the standard ``minimum'' ratio typically found in radiative shocks. This implies that there is no evidence for radiative shocks around the SNR. In known SMC SNRs, this ratio is typically found to be higher than 0.4 (\citealt{payne07}). 

\begin{figure}
\centering
\includegraphics[width=88mm]{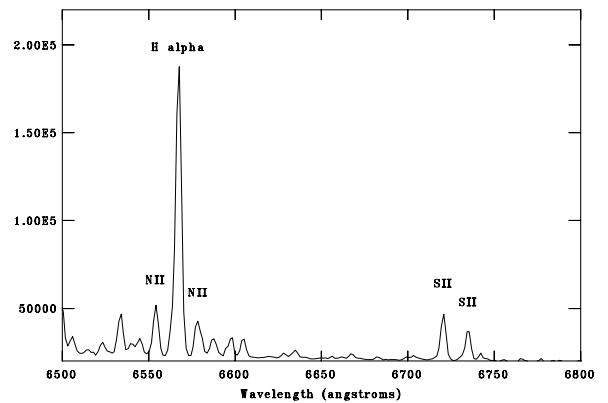}
\caption{WiFeS spectra of the central region of IKT~16.}
 \label{fig:ratio2}
\end{figure}

Similar results are found from analysis of the central region of IKT~16. Using QFitsView3\footnote{Written by Thomas Ott and freely available at www.mpe.mpg.de/~ott/dpuser/index.html}, a one-dimensional spectrum of the diffuse region excluding all visible stars was created (Fig.~\ref{fig:ratio2}). Analysis required the identification of cosmic rays on the 2-dimensional spectra while the IRAF task {\it splot} allowed determination of emission line counts based on gaussian fits. Standard errors were calculated based on an average rms noise of 600 counts; propagation of error methods were used for the addition or division of line fluxes. Using this method, we find a \Sii/\Ha\ ratio of 0.23 with a 2.1\% error. It is quite possible that there are very faint nebulae interspersed between the stars that do not meet the criteria for shocked regions. We cannot exclude the possibility of an \Hii region being present in the line of sight.


\subsection{Radio}
\label{sec:radio}
The available radio images show IKT~16 to be a very radio-weak SNR. There is evidence of faint diffuse radio structure throughout the remnant. The images are dominated by a source corresponding to the X-ray bright source, which appears to extend a distance of $\sim30\arcs$ towards the centre of the remnant. There is no evidence for a radio point source located within this emission. Table~\ref{table:radio} shows the integrated flux density measurements at frequencies from 843~MHz -- 8640 MHz. The overall radio spectrum (Fig.~\ref{fig:radiospec}) is well-described by a power-law with index $\alpha=-0.45\pm0.02$. The radio spectral index of the central source is found to be flatter ($-0.19\pm0.02$) than that of the SNR itself ($-0.97\pm0.35$). The SNR spectral index is unusually steep, especially given that IKT~16 is most likely an older (evolved) SNR, due to its rather large size. Usually, a steep gradient like this would suggest a much younger and more energetic SNR. However, in this case, the steepness can be contributed to missing short spacings at higher radio-continuum frequencies (4800 and 8640~MHz) and therefore missing flux. Specifically at 8640~MHz (where the ATCA primary beam is $\sim$300\arcsec) this SNR edges would be positioned close to the primary beam boundary where the flux tends to be significantly uncertain. We note that if we omit the two highest frequency data points from the SNR fit, the radio spectral index is --0.72, which is more in line with that found in other SNRs in the SMC (\citealt{filipovic05}).

\section{Discussion}
\label{sec:disc}

The only significant study of the X-ray properties of this remnant to date has been completed by VDH04. The \xmmn observations covering IKT~16 since publication of this paper have significantly improved the available statistics. This allows us to examine the properties of the SNR more accurately, as well as enabling study of the nature of the bright source located close to its centre. 

\subsection{Properties of the SNR}
\label{sec:discsnr}

In the analysis of IKT~16 conducted by VDH04, the bright source emission area was excluded from analysis, and the remaining emission was fit with both a single-temperature non-equilibrium ionization (NEI) model and a Sedov model. As we have shown, simply excluding the source emission area may not produce an accurate measure of the properties of the SNR, since $\sim25\%$ of the flux detected through this method will be from the point source. As this source is significantly harder than the SNR as a whole, this unduly influences the results obtained. Additionally, the previous study estimates the radius of the SNR to be $100\arcs$ compared with $128\arcs$ here. The new estimate is based on mosaicing of all available observations, through which the X-ray shock boundary is more clearly defined in all directions. This changes the estimate of the X-ray emitting volume, which also affects the derived Sedov parameters.

The Sedov properties obtained for both studies are shown in Table~\ref{table:sedov}. The properties we find justify our use of the Sedov approximation, as the swept-up mass significantly exceeds the ejecta mass but radiative cooling has not occurred yet. The parameters we derive are significantly different from those in VDH04. The temperature we find is cooler, which we attribute to the different methods of dealing with the point source mentioned in the previous paragraph. Along with our larger estimate of the radius of the remnant, this leads to a higher dynamical age, larger swept-up mass and lower initial explosion energy (as expected from Eq.\ref{eq:ne}-\ref{eq:energy}). 
In order to test whether this explains the difference in our results, we attempted to fit the SNR emission region with a Sedov model without accounting for spillover from the bright source. We find a good fit ($\chisq = 1099/1032$) with a higher shock temperature of $1.30\pm0.20$ keV. This is consistent with the value found by VDH04 ($1.76\pm0.65$ keV).  
VDH04 noted that the initial explosion energy appears high in comparison to typical SNRs, which they suggest is due to the remnant resulting from collapse of a massive star within a low-density circumstellar cavity. Our estimate of the explosion energy ($\sim10^{51}\ergsec$) is much more in line with that derived for other SNRs in the SMC. 

Compared with other SNRs in the SMC, IKT~16 is notably X-ray faint and is detected at higher electron temperature ($\sim1.0$ keV vs. $\sim0.2$ keV). While this is the largest confirmed SNR in the SMC there are several larger cousins in the Large Magellanic Cloud (LMC) \citep{2009SerAJ.179...55C,2010ApJ...725.2281K}. The radio emission from the SNR is also found to be faint (32 mJy at 1400~MHz; Table~\ref{table:radio}). The large size of the remnant, considering that it is still in the Sedov phase of evolution, suggest that the density of the SNR environment is low. This may also contribute to its X-ray and radio faintness. The low density implied is consistent with a SNR 
explosion inside a wind blown bubble, which are typically found inside molecular clouds. 
Such sites are associated with massive star formation, which would increase the likelihood 
of this system being a core-collapse SNR and the central source being associated with it. 
\citet{muller10} have carried out a survey of CO emission across the northern SMC, 
finding a series of molecular clouds near the position of IKT 16. Their presence strongly 
supports this scenario. The apparent X-ray faintness of the remnant may be additionally 
enhanced by high absorption attenuating the soft X-ray signal.  
Using a map of \hi density from \citet{stanimirovic99}, the total column density of the SMC in the direction of IKT~16 is $5.7\times10^{21}$cm$^{-2}$. The value we derive from spectral fitting is $\approx$60\% of this value, indicating that IKT 16 is located deep within the SMC. 
 
The soft X-ray image shows that the distribution of X-ray emission appears somewhat uneven inside the remnant, with more emission observed towards the northern shock boundary than to the south. 
The radio images also show more emission to the north of the remnant.  
The 3-colour X-ray image (Fig.~\ref{fig:ikt16images}, top left) suggests that the X-ray emission in the north may be at a higher temperature than in the rest of the remnant. This may be due to a combination of several factors, including fluctuations in the SMC absorption column, a higher concentration of SNR ejecta increasing the local metallicity or true variations in the plasma temperature of the ISM. Due to the limited X-ray statistics, it is not possible to carry out spatially resolved spectral analysis to investigate this further.


\begin{table}[t]
\caption{Physical properties of IKT~16 from best fit Sedov model.}
\centering
\begin{tabular}{lcccc}
\hline
Parameter  &  Units  &  This paper  &  VDH04  \\
\hline
Temperature  &  keV  &  1.03  &  1.76  \\
Radius  &  pc  &  37  &  29  \\
Electron Density  &  cm$^{-3}$  &  0.03  &  0.05/0.04$^{a}$  \\
Dynamical age  &  yr  &  14700   &  7500  \\
Swept-up mass  &  \Msun  &  232  &  124  \\
Explosion energy  &  $10^{51}$ erg  &  1.2  &  3.4  \\
\hline    
\end{tabular}
\\
$^{a}$ - VDH04 derive electron and proton densities separately, assumed to be approximately the same here.
\label{table:sedov}
\end{table}


\subsection{Nature of the Bright source}
\label{sec:discsrc}

The X-ray spectrum of the bright source is consistent with that of several different astrophysical objects, including X-ray binary systems, pulsar wind nebulae and background active galactic nuclei. Here, we attempt to discriminate between these possibilities. 

In fitting a power law model to the source, a significant absorption component is required in addition to foreground Galactic absorption. This suggests that it is not a Galactic object located along the line of sight. 

It is possible, however, that the source is a background AGN. The average power-law photon index of a type I AGN is measured to be $\sim1.9$ (\citealt{pounds90}), somewhat softer than what we see here. However, the absorption required in our spectral fit ($3.4\pm0.6\times10^{21}$ cm$^{-2}$) is significantly lower than that of the SMC along our line of sight, making this unlikely. We therefore attempted to fit a more complex model commonly found to fit AGN spectra: a power law plus a 0.1 keV black body component (\citealt{turner89}). Here, we find a reasonable fit with an extra absorption component of $3\pm1\times10^{21}$ cm$^{-2}$. The X-ray flux detected in this model is $\sim3\times10^{-13}\ergcms$.  

Studies of type~I AGN typically show the fluxes in X-ray (0.5--10 keV) and optical bands to lie within approximately an order of magnitude of each other (\citealt{maccacaro88}). For a source with this observed X-ray flux, we therefore expect there to be an optical counterpart at a V magnitude of between 14.8 and 20.8. We searched the Optical Gravitational Lensing Experiment (OGLE) (\citealt{udalski98}) and Magellanic Clouds Photometric Survey (\citealt{zaritsky02}) to try to find the optical counterpart. The closest match found to the X-ray position (searching down to a V band optical magnitude of 21.5) is at a distance of $2.5\arcs$, 2 times the combined errors in optical and X-ray positions. However, the two closest optical sources in the catalogue have optical colours consistent with stars within the SMC. The closest source with optical colours atypical for a main sequence star is at a distance of $3.2\arcs$, still within 3 sigma of our X-ray source position. This is a possible optical counterpart to our source.
 
We can also estimate the probability of finding a background AGN coincident with the SNR position. Studies of the log N-log S distribution of AGN (\eg \citealt{rosati02}) show that above our estimate of the unabsorbed X-ray flux of this object ($\sim4\times10^{-13}\ergcms$, 0.5-10 keV) we expect to find 3 AGN per square degree of the sky. The probability of finding one such source within 30\arcs of the centre of IKT 16 is 0.07\%. Although we are unable to rule out an AGN origin for this source, we conclude that it is statistically unlikely. We assume from this point that the source is located inside the SMC.       


As shown in \S\ref{sec:results}, the SNR shell is well-described by a circle of radius $128\arcs$. Presuming that the well-defined profile of the western shock boundary can be used to find the centre of the remnant, the bright source is located $\approx30\pm8\arcs$ (a transverse distance of $8\pm2$~pc) away from this point. Providing it is associated with the supernova event, it has traversed this distance in the dynamical age of the remnant (14700 yr), at an average speed of $580\pm100$ km s$^{-1}$. The direction of travel is not necessarily perpendicular to us, so this is a lower limit. However, unless the direction of travel is significantly out of this plane, the velocity of the object will not be much greater than this value. Asymmetry in SNR explosions typically gives the resulting compact object a random space velocity of $350-550$ km s$^{-1}$ (\citealt{lyne94}), so it is reasonable to suggest that the bright X-ray source is related to this object. There are two known objects which may produce the emission observed: a pulsar wind nebula or a microquasar.  

Pulsar wind nebulae (PWNe) are generated by dissipation of a pulsar's rotational energy through relativistic winds, producing synchrotron emission visible in radio to X-ray bands. For a review of the properties of these systems, see \citet{gaensler06}. $\sim50$ PWNe have been observed in the Galaxy and the LMC, and the majority are associated with known SNRs (``Composite'' SNR, \citealt{kaspi06}). The X-ray spectrum of known PWNe is well-described by a power-law of index ranging between $\Gamma\sim1.1-2.3$ (\citealt{cheng04}), consistent with the power law fit found here. The measured X-ray luminosity of $1.6\times10^{35}\ergsec$ is also similar to that of known PWNe in the Galaxy and the LMC (\eg \citealt{gaensler03}; \citealt{porquet03}; \citealt{slane04}). In our observations, the maximum extension of the X-ray source was found to be $6\arcs$, which corresponds to a radius of $\sim1$ pc at the distance of the SMC. \citet{gelfand09} find that in the spherically symmetric case, a pulsar wind nebula inside a Sedov-phase remnant at an age of 15~kyr should be undergoing compression due to the SNR reverse shock to a radius of $<1$ pc. This would be detected as a point source in \xmmn observations, consistent with our measurements. The radius measured in the \chandra observation is $0.3\pm0.2$ pc, consistent with this scenario. In this case, the movement of the source away from the centre of the SNR shell implies that spherical symmetry no longer holds, but as the distance travelled is only $\sim10\%$ of the shell diameter, this scenario may still be valid. Unfortunately, the limited spatial resolution of \xmmn and the distance to the SMC (60~kpc) preclude identification of a central pulsar, which would confirm the identity of this source as a pulsar wind nebula. The relatively low X-ray flux also prevents us from conducting timing analysis to search for pulsations. 
 
The ATCA images in Fig.\ref{fig:ikt16images} show the source to have a radio extent of $\sim40\arcs\times30\arcs$, elongated in the direction of the centre of the remnant. This extent is significantly greater than that observed in X-rays. This may be due to the longer cooling time for radio emission compared to that of X-ray emission. The radio emission may therefore trace the path of the nebula as it moves from the centre of the SNR, as is observed in ``relic'' PWN (\eg Vela X, \citealt{frail97}). The radio spectral index of this source is measured to be $\sim-0.2$, consistent with the typical radio spectrum found in PWNe (\citealt{gaensler06}). The absence of a radio point source in this emission rules out the presence of an established radio pulsar. However, confirmed PWNe (\eg LMC~SNR~B0453-685 and LMC~SNR~N\,206; \citet{gaensler03b,2005ApJ...628..704W}; Crawford et al. (in prep)) have been found where the central pulsar is not detected in radio observations, so this does not discount a PWN origin.  
  
The other possibility we consider here is a microquasar travelling through the SMC, where the extended radio emission is due to optically thin synchrotron radiation in relativistic jets. The X-ray emission in this case would come from an X-ray binary (XRB) system with a power-law of index 1.5--2.2, with radio jets observed at a spectral index between --0.8 and --0.4 (\eg \citealt{fender06}; \citealt{hannikainen06}). The X-ray power-law slope is consistent with this scenario, but the radio spectrum is somewhat flatter than expected ($\alpha\sim-0.2$). If it is associated with the SNR, and providing the western shock boundary can be used to estimate the original site of the supernova explosion, the velocity we infer for this object ($580\pm100$ km s$^{-1}$) is very high compared with the average kick velocity for even Low-Mass X-ray Binary (LMXB) systems ($180\pm80$ km s$^{-1}$; \citealt{brandt95}). In fact, the majority of known microquasars are High-Mass X-ray Binary (HMXB) systems, for which the corresponding typical kick velocity is 50 km s$^{-1}$ (also from \citet{brandt95}).  It is possible that the object may be within the SMC but not associated with the SNR, in which case this velocity measurement is not applicable. The relative stability of the X-ray flux between observations is somewhat unexpected, as many X-ray binary systems will exhibit state changes on a timescale of 10 years. From the available X-ray and radio evidence, we conclude that it is unlikely that the bright source is a microquasar. We note that no LMXB systems are currently known in the SMC, and they are very rare in the Galaxy. This also decreases the likelihood of this scenario. 
     
On the basis of current observations, it is most likely that the source is a PWN. However, we cannot entirely rule out a microquasar or AGN. A future deep \chandra observation will be able to determine the X-ray extent of this source accurately, and will thus conclusively determine its nature.     

\section{Conclusions}
\label{sec:conc}

We have carried out analysis of the spatial and spectral properties of the SMC SNR IKT~16, using archive and proprietary data from {\it XMM-Newton}, radio data from ATCA and optical images from MCELS. The properties of the SNR are consistent with it being in the Sedov-adiabatic phase of evolution. The dynamic age of the remnant and initial explosion energy are typical of the population of SNRs in the SMC. 

We find the X-ray point source near the centre of IKT~16 to be non-thermal, with an X-ray spectral index of $\sim1.6$. This source is coincident with a radio source which extends a distance of $\sim30\arcs$ in the direction of the centre of the SNR. We conclude that the X-ray source is unlikely to be a background AGN, due to the small probability of spatial coincidence with the SNR and the absence of an obvious optical counterpart. Two astrophysical objects are considered which may explain the X-ray and radio emission observed: a pulsar wind nebula or microquasar. The flatness of the radio spectrum, the one-sided morphology of the radio emission and inferred velocity of the source strongly favour a PWN origin, which may be confirmed by a future \chandra observation. If this is the case, IKT 16 is the second largest such system (after the Vela SNR) found to date.   
 

\begin{acknowledgements}
The \xmm\ project is supported by the Bundesministerium f\"ur Wirtschaft und Technologie/Deutsches Zentrum f\"ur Luft- und Raumfahrt (BMWI/DLR, FKZ 50 OX 0001) and the Max-Planck Society. RS acknowledges support from the BMWI/DLR grant FKZ 50 OR 0907. SM and AT acknowledge financial contribution from the agreement ASI-INAF I/009/10/0. We used the Karma/MIRIAD software packages developed by the ATNF. The Australia Telescope Compact Array is part of the Australia Telescope which is funded by the Commonwealth of Australia for operation as a National Facility managed by the CSIRO. We thank the Magellanic Clouds Emission Line Survey (MCELS) team for access to the optical images. 
We thank the anonymous referee for their comments, which have helped to improve this manuscript significantly.
\end{acknowledgements}

\bibliographystyle{aa}
\bibliography{references}

\end{document}